\documentclass[reqno,11pt]{amsart}

\usepackage{amsmath}
\usepackage{amsfonts}
\usepackage{indentfirst}
\usepackage{amssymb}

\newcommand{\bibit}[1]{\bibitem[#1]{#1}}
\newcommand{\paper}[1]{{\it #1}, }
\newcommand{\journal}[4]{#1 {\bf #2}, #3 (#4)}

\newcommand{\JPC}{J.\ Phys.\ C}

\newcommand{\PR}{Phys.\ Rev.}

\newcommand{\PRL}{Phys.\ Rev.\ Lett.}

\numberwithin{equation}{section}
\newtheorem{theorem}{Theorem}[section]

\newtheorem{lemma}[theorem]{Lemma}

\newcounter{eqs}

\renewcommand{\leq}{\;\leqslant\;}
\renewcommand{\geq}{\;\geqslant\;}

\newcommand{\upchi}{\raise1pt\hbox{$\chi$}}

\begin{document}

\vspace{0.5cm}

\title[Existence of Solutions to the Bethe Ansatz Equations: 1D Hubbard Model]{Existence of Solutions 
to the Bethe Ansatz Equations for the 1D Hubbard 
Model: Finite Lattice and Thermodynamic Limit}

\author{Pedro S. Goldbaum}

\maketitle

\vspace{-0.8cm}
\begin{center}
{\small Department of Physics, Princeton University \\
Jadwin Hall, Princeton, NJ 08544, USA.}
\end{center}

\vspace{0.5cm}

\renewcommand{\thefootnote}{}
\footnote{E-mail: {\rm goldbaum@princeton.edu}.}

\begin{quote}
{\small
{\bf Abstract.}
In this work, we present a proof of the existence of real and ordered solutions to the generalized 
Bethe Ansatz equations for the one 
dimensional Hubbard model on a finite lattice, with periodic boundary conditions. The existence of 
a continuous set of solutions extending from any $U>0$ to $U=\infty$ is also shown. This 
continuity property, when combined with the proof that the wavefunction obtained with 
the generalized Bethe Ansatz is normalizable, is relevant to the question of whether or not the solution gives us the ground 
state of the finite system, as suggested by Lieb and Wu. Lastly, for the absolute ground state at half-filling, 
we show that the solution converges to a distribution in the thermodynamic limit. This limit distribution satisfies the 
integral equations that led to the well known solution of the 1D Hubbard model.
}

\vspace{1mm}

\end{quote}

\section{Introduction}
In 1968, Lieb and Wu \cite{LW1} solved the one-dimensional Hubbard model \cite{H}, and showed the absence of Mott 
transition in the ground state. A new version of this result \cite{LW2}, with additional details 
on the exact solution for the ground state, was recently published, motivated by the increasing importance of the 
1D Hubbard model in Condensed Matter Physics.

Despite being one of the most studied models to describe interacting electron systems, 
many questions concerning the Hubbard model remain as open problems, and the 1D Hubbard model is the only one for 
which an exact solution was found. Still, many results concerning the Bethe Ansatz to the Hubbard model 
rely on unproved hypotheses. The purpose of this paper is to answer some of the questions raised 
in the recent paper by Lieb and Wu, concerning the existence of a solution to the Bethe Ansatz 
equations for the 1D Hubbard model on a finite lattice, and whether or not this solution gives 
us the ground state. We will also discuss the solution in the thermodynamic limit.

Consider a system with $N$ electrons on a lattice with $N_a$ sites. The one-dimensional Hubbard 
model is described by the hamiltonian 
\begin{equation}
\nonumber
H=-\sum_{i=1}^{N_a}\sum_{\sigma}(c^{\dagger}_{i+1,\sigma}c_{i,\sigma}+h.c.)+U\sum_{i=1}^{N_a}n_{i\uparrow}
n_{i\downarrow},
\end{equation}
where $c^{\dagger}_{i,\sigma}$, $c_{i,\sigma}$ are the fermion creation and annihilation operators 
for an electron with spin $\sigma=\uparrow$ or $\downarrow$, at site $i$, and $n_{i\uparrow}$, 
$n_{i\downarrow}$ are the corresponding occupation number operators. We are using here periodic 
boundary conditions ($N_a+1 \equiv 1$), so we are actually working on a ring with 
$N_a$ sites. Particle-hole symmetry permits us to restrict to the case $N \leq N_a$.

Since the hamiltonian commutes with $\sum_{i=1}^{N_a}n_{i\uparrow}$ and $\sum_{i=1}^{N_a}n_{i\downarrow}$, 
we can look for energy eigenstate with a fixed number of spins up and down. Let us denote these states 
by $|M,M'>$, where $M$ is the number of spin-down, and $M'$ the number of spin-up electrons. Spin-up and spin-down 
symmetry allows us to restrict to the case $M \leq M'$. 

We can expand $|M,M'>$ in a basis of states of localized electrons,
\begin{equation}
\label{state}
|M,M'>=\sum_{1\leq x_i \leq N_a}f(x_1,\dots ,x_N)|x_1,\dots ,x_N>,
\end{equation}
where $|x_1,\dots ,x_N>$ denotes the state where the spin-down electrons occupy the sites $x_1,\dots,
x_M$, and the spin-up electrons are located at $x_{M+1},\dots ,x_N$.

Our states are defined in the region $R=\{1 \leq x_i \leq N_a \, , \, i=1,\dots ,N \}$. Given any permutation 
$Q:\{1,\dots,N\} \rightarrow \{Q1, \dots, QN \}$, we can define the region
\begin{equation}
\nonumber
R_Q=\{1 \leq x_{Q1} \leq x_{Q2} \leq \dots \leq x_{QN} \leq N_a \} \subset R
\end{equation}
It is clear that 
\begin{equation}
\nonumber
\cup_Q R_Q = R,
\end{equation}
where the union is taken over the $N!$ possible permutations. We should always keep in mind that the
$x_1,\dots, x_M$ denote the positions of the spin-down electrons.

The solution of the 1D Hubbard model is based on a generalized version of the Bethe-Ansatz \cite{Be}. For a sequence of 
unequal and ordered numbers $k_1 < k_2 < \dots < k_N$, we assume that the energy eigenstates are 
given by \eqref{state}, where the amplitudes $f(x_1,\dots, x_N)$ are given in the region $R_Q$ by
\begin{equation}
\label{coef}
f(x_1,\dots ,x_N)=\sum_P A_Q\{P\}\exp{\bigl[ i\sum_{j=1}^Nk_{P_j}x_j \bigr]}.
\end{equation}
The $(N!)^2$ coefficients $A_Q\{P\}$ have to be determined in order for the state to be well defined 
in the boundary of two regions $R_Q$ and $R_{Q'}$. Also, we should make sure that the resulting state is 
antisymmetric with respect to exchange of two identical particles, and that it satisfies the periodic 
boundary conditions.

Those restrictions result in a system of $(N!)^2$ linear and homogeneous equations for the coefficients 
$A_Q\{P\}$. Fortunately, these equation were studied in detail by Gaudin \cite{G} and Yang \cite{Y}, in the context of 
a system of fermions with delta interaction. The only difference here is that we are working on a 
lattice, and not in the continuum, so the $k_j$ that appear on the system of equations have to be 
substituted by $\sin{k_j}$, which does not interfere with the algebraic analysis. 

The condition obtained in \cite{G} and \cite{Y} for the existence of a non-trivial solution 
for the coefficients $A_Q\{P\}$ can be written as the generalized Bethe-Ansatz equations
\begin{eqnarray}
\label{ba1}
& & e^{ik_jN_a}=\prod_{\beta=1}^M\frac{i\sin{k_j}-i\Lambda_\beta-U/4}{i\sin{k_j}-i\Lambda_\beta+U/4}
\quad , \quad j=1,\dots,N \quad , \\
\label{ba2}
& & \prod_{j=1}^N\frac{i\sin{k_j}-i\Lambda_\alpha-U/4}{i\sin{k_j}-i\Lambda_\alpha+U/4}=
-\prod_{\beta=1}^M\frac{-i\Lambda_\beta+i\Lambda_\alpha+U/2}{-i\Lambda_\beta+i\Lambda_\alpha-U/2}
\quad , \quad \alpha=1,\dots,M \quad ,
\end{eqnarray}
in terms of our original set $\{k_j\}_{j=1}^N$, and a set of auxiliary parameters $\{\Lambda_\alpha
\}_{\alpha=1}^M$.

Defining
\begin{equation}
\nonumber
\theta(x)=-2\tan^{-1}(2x/U)
\end{equation}
and taking the logarithm of the equations above, we obtain the Lieb-Wu equations
\begin{eqnarray}
\label{bethe1}
N_a k_j=2\pi I_j+\sum_{\beta} \theta (2\sin k_j-2\Lambda_{\beta}), \quad j=1,\dots,N \ ,\\
\sum_j \theta (2\sin k_j-2\Lambda_{\alpha})=2\pi J_{\alpha}-\sum_{\beta} \theta(\Lambda_{\alpha}-\Lambda_{\beta}),
 \quad \alpha=1,\dots, M \ .
\label{bethe2}
\end{eqnarray}
The coefficients $I_j$ are integers if $M$ is even, and half-integers otherwise. Similarly, $J_\alpha$ are integers 
if $N-M=M'$ is odd and half-integers otherwise. We will restrict to the case where $M$ is odd, 
and $N$ is even, since in this case the ground state is unique for every $U$. For the 
ground state, we can choose the $I_j$ and $J_\alpha$ that will give us the correct solution in the 
limit $U\rightarrow \infty$. In this limit, the equations decouple, and we have
\begin{equation}
I_j=j-\frac{N+1}{2} \quad , \quad J_\alpha=\alpha-\frac{M+1}{2}.
\nonumber
\end{equation}

In the thermodynamic limit, taking $N_a,N,M,M'\rightarrow \infty$, keeping their ratios 
fixed, \eqref{bethe1} and \eqref{bethe2} become integral equations. The analysis of the resulting equations 
led to the well known solution of the 1D Hubbard model.

Some questions concerning the finite lattice case remained as open problems. The purpose of this 
paper is to provide rigorous results to some of those problems. In particular, we are going to show that the 
equations \eqref{bethe1}, \eqref{bethe2} do indeed have a solution, and that it is ordered in $j$ and $\alpha$ 
($k_{j+1}>k_j$, $\Lambda_{\alpha+1}>\Lambda_{\alpha}$). The ordering is important in the derivation of the 
integral equations in the thermodynamic limit. We also show that there is a continuous curve of 
solutions extending from $U=\infty$ to any $U>0$. We are ready to state our main result:

\begin{theorem}
The equations \eqref{bethe1} and \eqref{bethe2} have at least one real solution satisfying 
\begin{eqnarray}
 -\pi \leq k_1 < & \cdots &  < k_N \leq \pi, \nonumber \\
\Lambda_1 < & \cdots & < \Lambda_M. \nonumber
\end{eqnarray}
Furthermore, there is a continuous curve of solutions defined by $\gamma(t):[0,1] \rightarrow 
\mathbb{R}^{N+M+1}$, where
\begin{equation}
\nonumber
\gamma(t)=(k_1(t),\dots, k_n(t), \Lambda_1(t), \dots, \Lambda_M(t),U(t)),
\end{equation}
extending from any $U>0$ to $U=\infty$.
\end{theorem}

We know that the state obtained by the solution to \eqref{bethe1}, \eqref{bethe2} correspond to the true ground state 
in the limit $U=\infty$, since we chose the coefficients $I_j$ and $J_\alpha$ in order to match 
these states. To prove that the ground state is given by the Bethe Ansatz solution for any 
positive $U$, we need a few additional facts. Since $M$ and $M'$ are odd, the ground state is non-degenerate for 
any $U$ \cite{LW2}, and our result shows that 
starting with the solution at $U=\infty$, and going along the curve $\gamma(t)$, the state is 
indeed the ground state, provided $f(x_1,\dots,x_N)$ defined by  
\eqref{coef} does not vanish for all $(x_1,\dots,x_N)$.

On section 4, we prove that the generalized Bethe Ansatz equations have a solution that is 
algebraic in $U$. Since the Bethe Ansatz wavefunction is a rational function in the variables $z_j=e^{ik_j}$, 
$\Lambda_\alpha$ and $U$ (see \cite{W}), its norm will have at most finitely many zeros and poles as a function of $U$. We 
show that we can redefine 
our wavefunction at these problematic points, in order to obtain the physical solution for all $U$.

Once the existence of a solution  that corresponds to the ground 
state of our finite system is shown, our next goal is to determine what happens with the set of $k$-s and $\Lambda$-s in 
the thermodynamic limit. As we increase the lattice size, a sequence of distributions measuring the 
density
 of $k$-s and $\Lambda$-s can be defined. In the thermodynamic limit, we show that a subsequence of these distributions converge to
the solution of the integral equations derived by Lieb and Wu.

We are particularly interested in the absolute ground state at half-filling, in which case an exact solution 
can be found, including an explicit expression for the ground state energy. In this case, we show that 
 the whole sequence converges to the explicit solution obtained in \cite{LW1}.

\section{Existence of a solution}

The goal in this section is to prove the existence of solutions to the Lieb-Wu equations that are 
ordered in $j$ and $\alpha$ ($k_{j+1}>k_j$, $\Lambda_{\alpha+1}>\Lambda_{\alpha}$). The proof will follow 
from Brouwer's fixed point theorem, but first we need to define a convenient map to apply the theorem.

\subsection{Defining the map}

Because of the symmetry of the coefficients $I_j$ and $J_\alpha$, we can look for symmetric solutions 
of the form $k_j=-k_{N-j+1}$, and $\Lambda_{\alpha}=-\Lambda_{M-\alpha+1}$. Therefore, our set of $k$-s 
and $\Lambda$-s are totally defined by 
\begin{eqnarray}
\mathbf{k} & = & \{ k_j \} \quad , \quad j=\frac{N}{2}+1,\dots ,N \quad ,\nonumber\\
\mathbf{\Lambda} & = & \{ \Lambda_\alpha \} \quad , \quad \alpha=\frac{M+3}{2}, \dots ,M \quad \nonumber.
\end{eqnarray}
The other components of $\mathbf{k}$ and $\mathbf{\Lambda}$, are obtained by the symmetry condition. In 
particular, $\Lambda_{(M+1)/2}=0$.

We can define the map $\phi:{\mathbb{R}}^{\frac{N}{2}+\frac{M-1}{2}}\rightarrow {\mathbb{R}}^{\frac{N}{2}+\frac{M-1}{2}}$ by

\begin{equation}
\nonumber
\phi(\mathbf{k}, \mathbf{\Lambda})=(\mathbf{k'},\mathbf{\Lambda'}).
\end{equation}
where $\mathbf{\Lambda'}$ is defined by

\begin{equation}
\label{deflambd}
\sum_j \theta (2\sin k_j-2\Lambda'_{\alpha})=2\pi J_{\alpha}-\sum_{\beta} \theta(\Lambda_{\alpha}-\Lambda_{\beta}), 
\end{equation}
for $(M+3)/2 \leq \alpha \leq M$, and $\mathbf{k'}$ by

\begin{equation}
k'_{\frac{N}{2} + 1}= \max\{0, \frac{1}{N_a}(2\pi I_{\frac{N}{2}+1}+\sum_{\beta} \theta (2\sin k_{\frac{N}{2}+1}
-2\Lambda_{\beta}))\},
\end{equation}
whereas for $i>\frac{N}{2}+1$

\begin{equation}
k'_i=\max\{k'_{i-1}, \frac{1}{N_a}(2\pi I_i+\sum_{\beta} \theta (2\sin k_i-2\Lambda_{\beta}))\}.
\end{equation}

The function $\mathbf{\Lambda'}(\mathbf{k},\mathbf{\Lambda})$ is well defined since the left side 
of \eqref{deflambd} is continuous and strictly increasing in $\Lambda'_{\alpha}$. It is also clear 
that $\mathbf{k'}(\mathbf{k},\mathbf{\Lambda})$ is continuous, so the map $\phi(\mathbf{k}, \mathbf{\Lambda})
=(\mathbf{k'},\mathbf{\Lambda'})$ is continuous. 

It is also important to observe that
\begin{equation}
\nonumber
\Lambda'_\alpha < C_{N,U},
\end{equation}
where $C_{N,U}$ is a constant for each value of $N$ and $U$. Indeed, from \eqref{deflambd} we 
have
\begin{equation}
\nonumber
\sum_j \theta (2\sin k_j-2\Lambda'_{\alpha}) \leq \frac{(M-1)}{2}2\pi+M\pi \leq (N-1)\pi,
\end{equation}
which implies
\begin{equation}
\nonumber
\frac{4\Lambda'_\alpha}{U} < \tan{\Bigl(\frac{\pi}{2}-\frac{\pi}{2N}\Bigr)} +\frac{4}{U},
\end{equation}
or
\begin{equation}
\nonumber
\Lambda'_\alpha < \frac{U}{4}\tan{\Bigl(\frac{\pi}{2}-\frac{\pi}{2N}\Bigr)}+1 \equiv C_{N,U}.
\end{equation}

\subsection{The fixed point theorem} Now we are ready to state the result 
that allows us to use Brouwer's fixed point theorem.
\begin{lemma}
If we define the domain $\Omega \subset \mathbb{R}^{\frac{N}{2}+\frac{M-1}{2}}$ by
\begin{eqnarray}
0 \leq k_{\frac{N}{2}+1} \leq & \dots & \leq k_N \leq \pi, \nonumber \\
0 \leq \Lambda_{\frac{M+3}{2}} \leq & \dots &  \leq \Lambda_M \leq C_{N,U}, \nonumber
\end{eqnarray}
we have
\begin{equation}
\nonumber
\phi (\Omega) \subset \Omega.
\end{equation}
\end{lemma}
{\bf{Proof.}} By definition,
\begin{equation}
\nonumber
0 \leq k'_{\frac{N}{2}+1} \leq \dots \leq k'_N.
\end{equation}
Also,
\begin{equation}
\nonumber
k_N \leq \frac{1}{N_a}\Bigl[2\pi\Bigl(\frac{N-1}{2}\Bigr)\Bigr] < \pi \frac{N}{N_a} \leq \pi.
\end{equation}

Regarding $\Lambda_\alpha$, we have
\begin{equation}
\nonumber
\sum_j \theta (2\sin k_j-2\Lambda'_{\alpha}) >0 \quad \Rightarrow \quad \Lambda'_\alpha > 0.
\end{equation}
Furthermore, the right side of \eqref{deflambd} is increasing in $\Lambda_\alpha$ and 
$J_{\alpha+1} > J_\alpha$, whereas the left side is increasing in $\Lambda'_\alpha$. Restricting 
the map $\phi$ to $\Omega$, this implies
\begin{equation}
\nonumber
\Lambda'_{\alpha +1} \geq \Lambda'_\alpha.
\end{equation}
Combining these results with $\Lambda'_M < C_{N,U}$ we conclude the proof of the lemma.

Since $\Omega$ is compact and convex (in particular, it has no ``holes''), there is a continuous and 
invertible map 
\begin{equation}
\nonumber
\psi:\Omega \rightarrow D^{\frac{N}{2}+\frac{M-1}{2}},
\end{equation}
where $D^{\frac{N}{2}+\frac{M-1}{2}}$ is a disk in $(N+M-1)/2$ dimensions. The map 
\begin{equation}
\nonumber
\psi \circ \phi \circ \psi^{-1}:D^{\frac{N}{2}+\frac{M-1}{2}}\rightarrow D^{\frac{N}{2}+\frac{M-1}{2}}
\end{equation}
is continuous and takes the disk into the disk, and by Brouwer's fixed point theorem, it must have a fixed point. Since 
$\psi$ is one-to-one, the fixed point of $\psi$ corresponds to a fixed point of $\phi$.

\subsection{The fixed point is a solution to the Bethe Ansatz equations} Since we artificially 
introduced the maximum in the definiton of $\mathbf{k'}$, we need to show that the fixed point
$(\mathbf{k^*},\mathbf{\Lambda^*})$, such that
\begin{equation}
\nonumber
\phi(\mathbf{k^*},\mathbf{\Lambda^*})=(\mathbf{k^*},\mathbf{\Lambda^*}),
\end{equation}
is indeed a solution to the original equations.

In other words, we need to show that 
\begin{equation}
\nonumber
0 < k^*_{\frac{N}{2}+1} < \dots < k^*_N.
\end{equation}

Taking $k_j=0$, we have
\begin{equation}
\nonumber
k'_j=\frac{2\pi I_j}{N_a} > 0 \quad \Rightarrow \quad k^*_{\frac{N}{2}+1}>0.
\end{equation}
That also implies
\begin{equation}
\nonumber
k^*_{\frac{N}{2}+1}=\frac{1}{N_a}\bigl[2\pi I_{\frac{N}{2}+1}+\sum_{\beta} \theta 
(2\sin k^*_{\frac{N}{2}+1}-2\Lambda_{\beta})\bigr].
\end{equation}
Let us consider now 
\begin{equation}
\nonumber
k^*_{\frac{N}{2}+3}=\max \bigr\{ k^*_{\frac{N}{2}+1},\frac{1}{N_a}\bigl[2\pi I_{\frac{N}{2}+3}+\sum_{\beta} \theta 
(2\sin k^*_{\frac{N}{2}+3}-2\Lambda_{\beta})\bigr] \bigl\}.
\end{equation}
Assuming that $k^*_{\frac{N}{2}+3}=k^*_{\frac{N}{2}+1}$, we have
\begin{equation}
\nonumber
k^*_{\frac{N}{2}+3}=\max \bigl\{ k^*_{\frac{N}{2}+1}, k^*_{\frac{N}{2}+1}+\frac{2\pi}{N_a} \bigr\}
>k^*_{\frac{N}{2}+1},
\end{equation}
which is a contradiction. Therefore, $k^*_{\frac{N}{2}+3} > k^*_{\frac{N}{2}+1}$, and proceeding 
by induction we get
\begin{equation}
\nonumber
k^*_{\frac{N}{2}+1} < k^*_{\frac{N}{2}+3} < \dots < k^*_N.
\end{equation}

We have therefore concluded the proof that the fixed point $(\mathbf{k^*},\mathbf{\Lambda^*})$ 
satisfies \eqref{bethe1} and \eqref{bethe2}.

\section{Existence of a continuous curve of solutions}

We want to analyze now what happens to our solutions as $U$ changes. Our goal is to show that 
there is a connected set of solutions extending from any $U>0$ to $U=\infty$.

From what we have seen in the last section, we know that on the part of the boundary of $\Omega$ defined by 
$k_{\frac{N}{2}+1}=0$ or $k_{i+1}=k_i$, 
the vector field defined by 
\begin{equation}
\nonumber
\mathbf{v}_U(\mathbf{k},\mathbf{\Lambda})=\phi_U(\mathbf{k},\mathbf{\Lambda})-(\mathbf{k},\mathbf{\Lambda})
\end{equation}
has a normal component that always points inward. Also, for $\Lambda_\alpha=0$ we have $\Lambda'_\alpha>0$, 
and $\Lambda_\alpha=\Lambda_{\alpha+1}$ implies $\Lambda'_\alpha<\Lambda'_{\alpha+1}$. Therefore, the vector 
field actually points inward in all the boundary of $\Omega$.

All we need to prove then is that given any family of continuous vector fields
\begin{equation}
\nonumber
\mathbf{v}_t:D^n \rightarrow \mathbb{R}^n \quad , \quad t \in [0,1],
\end{equation}
such that the normal component of $\mathbf{v}_t$ points inward on the boundary of the disk $D^n$, there is 
a connected subset of $D^n \times [0,1]$ in which $\mathbf{v}_t(\mathbf{k},\mathbf{\Lambda})=0$, extending 
all the way from $t=0$ to $t=1$.

The proof of this statement is trivial if for each $t \in [0,1]$ the vector field $\mathbf{v}$ is non-degenerate.
 (Here, we consider a vector field to be non-degenerate if, for every point $x$ such that $\mathbf{v}(x)=0$, the 
determinant of the jacobian $|\partial_j v_i|$ does not vanish.) In this case, since the zeros of the vector field are always isolated, each zero at $t=0$ will follow 
a smooth curve as we increase $t$ from $0$ to $1$, by the implicit funtion theorem.

In the general case, however, zeros can collide with each other and disappear, as we change $t$, 
so the proof is not that simple. Let us assume now that we have a general family of vector 
fields, possibly degenerate. 

We can always approximate $\mathbf{v}_t$ by a polynomial vector field, in the sense that for 
every $\epsilon>0$ there is a polynomial vector field $\mathbf{p}$ such that $\| \mathbf{v}_t-\mathbf{p}\| < \epsilon$. 
If $\mathbf{v}_t$ is a polynomial, 
it will have a finite number of sets of connected zeros. We can define an 
index for every isolated set of zeros. For an isolated zero $\mathbf{x}_0$, we define its index \cite{GP}
by the degree of the map $\phi:S_\epsilon \rightarrow S^{n-1}$ defined by
\begin{equation}
\nonumber
\phi(\mathbf{x})=\frac{\mathbf{v}_t}{|\mathbf{v}_t|},
\end{equation}
where $S_\epsilon$ is a sphere center at $\mathbf{x}_0$ with radius $\epsilon$, and $\epsilon$ is sufficiently small 
so that $\mathbf{x}_0$ is the only zero of $\mathbf{v}_t$ in its interior. 
So the index is just a measure of how many times we map the sphere $S_\epsilon$ 
into the sphere $S^{n-1}$.

For any connected set of zeros, we can define the index analogously, by isolating the set from other 
zeros by a surface that can be smoothly deformed into a sphere. Alternatively, we could define the index by 
adding a small perturbation to remove the degeneracy, and adding the index of all resulting isolated 
zeros. A general definition of the index for vector fields can be found in \cite{GS}. 

Since $\mathbf{v}_t$ always points inwards in the boundary of the disk, the sum of the indices of 
all connected sets of zeros is equal to one, for all $t$. As $t$ varies from $0$ to $1$, the zeros 
describe a continuous trajectory. A given zero of index +1 cannot simply disappear, unless it collides 
with a zero of index $-1$. Every connected set of zeros can at some point break up into other sets, 
provided the sum of indices is preserved. Also, new sets of zeros can be created, provided their total 
index is 0.

Let $Z \subset D^n \times [0,1]$ be the set of all zeros of $\mathbf{v}$, where we now drop the index $t$, and consider 
the vector field defined on the cylinder $D^n \times [0,1]$. If a zero of $\mathbf{v}_0$ with index +1 is not connected to a zero of 
$\mathbf{v}_1$ by Z, 
then it has to be connected to another zero of $\mathbf{v}_0$ with index $-1$.
Since the total index is always equal to 1, there are not enough zeros of index -1 to annihilate all the +1 zeros. 
and there will be at least one connected set of zeros from 
$t=0$ to $t=1$.

We can find a sequence of vector fields $\{\mathbf{w}_n\}$, such that $\mathbf{v}+\mathbf{w}_n$ is 
a polynomial vector field and $\lim_{n\rightarrow \infty}|\mathbf{w}_n|=0$. For each $n$, we have a continuous curve 
of zeros of $\mathbf{v}+\mathbf{w}_n$, from $t=0$ to $t=1$. What we need to do is to prove that the curve cannot be 
disrupted as we take the limit $|\mathbf{w}_n| \rightarrow 0$.

Let us also define the set
\begin{equation}
Z'=D_0 \cup Z \cup D_1
\nonumber
\end{equation}
where $D_0$ is the disk at $t=0$, and $D_1$ the disk at $t=1$. Consider the topology $\tau^*$ in 
$Z'$ defined by taking the intersection of the open sets in the metric topology $\tau$ of $D \times [0,1]$ 
with $Z'$.

Let us assume for a moment that there is no subset of $Z$ connecting $D_0$ to $D_1$. In this case, we
can find two disjoint open sets $X_0,X_1 \in \tau^*$ such that
\begin{equation}
\nonumber
D_0 \subset X_0 \quad , \quad D_1 \subset X_1,
\end{equation}
and $X_0\cup X_1=Z'$. 
But $X_0$ and $X_1$ are closed sets in the metric topology $\tau$ (defined by the open balls in the usual 
metric on the cylinder), since they are given by zeros of 
a continuous vector field. Therefore we can find disjoint open sets $O_1,O_2 \in \tau$ such that
\begin{equation}
\nonumber
X_0 \subset O_0 \quad , \quad X_1 \subset O_1.
\end{equation}

But the set $(O_1\cap O_2)^c$ is nonempty and closed, so 
\begin{equation}
|\mathbf{v}(x)|\neq 0 \quad , \quad \forall x \in (O_1\cap O_2)^c,
\nonumber
\end{equation}
implies
\begin{equation}
|\mathbf{v}(x)| > \delta \quad , \quad \forall x \in (O_1\cap O_2)^c.
\nonumber
\end{equation}
Hence, for $|\mathbf{w}|$ sufficiently small, 
\begin{equation}
|(\mathbf{v}+\mathbf{w})(x)| > \delta/2,
\nonumber
\end{equation}
and there is no curve of zeros of $\mathbf{v}+\mathbf{w}$ connecting $D_0$ to $D_1$. This is in 
contradiction with the existence of a sequence of polynomial vector fields converging to 
$\mathbf{v}$.

Therefore, we conclude the proof of the existence of a connected set of solutions of the Bethe-Ansatz 
equations, extending from any $U>0$ to arbitrarily large values. Along this curve, the energy given by 
\begin{equation}
\nonumber
E(U(t))=-2\sum_{j=1}^N\cos{k_j(t)}
\end{equation}
is continuous in $t$.

\section{Non-vanishing norm of the wavefunction}
As we discussed before, the existence of a continuous curve of solutions implies that 
the right side of \eqref{state} 
is indeed the true ground state, provided that its norm is not zero. However, since we are working on a lattice, and not in the continuum, the wavefunction \eqref{state}
 could in principle vanish ($f(x_1,\dots,x_N)\equiv 0$)
, even though the coefficients $A_Q\{P\}$ are not all zero. Therefore, to complete the proof 
that \eqref{state} is the true ground state, we need to prove that $f(x_1,\dots,x_N)$ 
does not vanish identically. 

We will follow here the same strategy used by C.\ N.\ Yang and C.\ P.\ Yang \cite{YY} to prove that the Bethe Ansatz 
does indeed give the true ground state for the anisotropic Heisenberg model. 

Let us first define the variables $z_j$ by

\begin{equation}
\nonumber
z_j=\left \{
\begin{array}{ll}
e^{ik_j} \quad , \quad j=1,\dots,N \nonumber ,\\
\Lambda_{j-N} \quad , \quad j=N+1,\dots,N+M .\nonumber
\end{array}
\right .
\end{equation}

We can write the equations \eqref{ba1}, \eqref{ba2} 
as polynomial equations in the variables $z_j$ and $U$. So we have a system of $N+M$ 
polynomial equations, in $N+M+1$ variables,
\begin{equation}
p_i(z_j,U)=0 \quad , \quad i=1,\dots,N+M,
\label{baalgebraic}
\end{equation}
and our goal is to find solutions given by algebraic functions $z_j(U)$.

Let us remind ourselves of some basic facts concerning polynomial equations in several variables. A more detailed 
background on the subject can be found in \cite{CLO}. 
Consider the space of polynomials in $n$ complex variables, with complex coefficients, which we denote by $\mathbb{C}[x_1,\dots,x_n]$. Given 
a set of polynomials $\{f_i\} \subset \mathbb{C}[x_1,\dots,x_n]$, where $i=1,\dots,m$, we can define an ideal 
$I(f_1,\dots,f_m) \subset \mathbb{C}[x_1,\dots,x_n]$ by
\begin{equation}
\nonumber
I(f_1,\dots,f_m)=\{ \sum_{i=1}^{m}h_i f_i\ ,\ h_1,\dots ,h_m \in \mathbb{C}[x_1,\dots,x_n] \}.
\end{equation}
We can also define an affine variety $V(I)\subset \mathbb{C}^n$ by 
\begin{equation}
\nonumber
V(I)=\{(a_1,\dots,a_n)\in \mathbb{C}^n \ : \ f(a_1,\dots,a_n)=0\ , \ \forall f \in I\}.
\end{equation}
$V(I)$ is the set of solutions to the original polynomial equations 
\begin{equation}
\label{polequation}
f_1(x_1,\dots,x_n)=f_2(x_1,\dots,x_n)=\cdots=f_m(x_1,\dots,x_n)=0.
\end{equation}

We say that the set of polynomials $\{f_j\}$ form a basis to the ideal $I$. But an ideal can be 
defined by many bases. A particularly convenient basis for an ideal is the Groebner basis 
(see appendix A or \cite{CLO} for details). 
The Groebner basis is a set of polynomials with the same roots as the original set, with the nice 
property that it contains polynomials where some of the variables are eliminated. 
We will be using a Groebner basis for the ideal defined by the Bethe Ansatz equations shortly, but we still 
need to define the projection of an affine variety.

Given $V(I)$, we can define $\pi^{(k)}:V(I) \subset \mathbb{C}^n\rightarrow \mathbb{C}$ by
\begin{equation}
\nonumber
\pi^{(k)}(a_1,\dots,a_n)=a_k
\end{equation}
This projection gives us the the values of $x_k$ for which the system of polynomial equations has at 
least one solution.

We are ready to state the technical result that will assist us in our proof.

\begin{lemma}
Let $V \subset \mathbb{C}^n$  be an affine variety. Then one of the following has to be true:
\begin{itemize}
\item
$\pi^{(k)}(V)$ consists of finitely many points (possibly zero),
\item
$\mathbb{C}-\pi^{(k)}(V)$ consists of finitely many points (possibly zero). 
\end{itemize}
\end{lemma}
{\bf{Proof}}: If we define the ideal $I^{(k)}=I\cap \mathbb{C}[x_k]$, given by the polynomials in 
$I$ that depend only on the variable $x_k$, we have
\begin{equation}
\nonumber
V(I^{(k)})-W \subset \pi^{(k)}(V) \subset V(I^{(k)}),
\end{equation}
where $W$ is a proper subset of $V(I^{(k)})$. The right side of this relation is trivial, since 
$(a_1,\dots, a_k, \dots, a_n)\in V(I)$ implies that $a_k$ is a root of any polynomial in $I^{(k)}$. The 
left side follows from \cite{CLO} (Theorem 3, p.123).

But $V(I^{(k)})$ and $W$ are affine varieties in $\mathbb{C}$, which can either be $\mathbb{C}$ or a set with finitely many 
points (roots of a polynomial equation $p(x_k)=0$). If $V(I^{(k)})=\mathbb{C}$, $W$ is a finite set and 
$\mathbb{C}-\pi^{(k)}(V)$ consists of finitely many points. Otherwise, $\pi^{(k)}(V)$ is finite.

As a first application of this lemma, we see that our system of polynomial equations will actually have a solution for any complex $U$, except for 
at most finitely many values, since we proved that it has at 
least one solution for $U$ real and positive,. If we choose the ordering $z_1 > z_2 > \dots >z_{N+M} > U$, our Groebner 
basis will not have any polynomial depending only on $U$. Therefore, our last polynomial in the basis 
should be of the form
\begin{equation}
\nonumber
p(z_{M+N},U)=0.
\end{equation}
There is nothing special about the ordering we chose, so can also construct a Groebner basis with 
the ordering $z_1  > \dots >z_{k} > U$, for any $k$ between $1$ and $N+M$. Therefore, the solutions 
of \eqref{baalgebraic} will have to satisfy
\begin{eqnarray}
p^{(1)}(z_1,U) & = & 0 \nonumber ,\\
p^{(2)}(z_2,U) & = & 0 \nonumber ,\\
\vdots \nonumber \\
p^{(M+N)}(z_{M+N},U) & = & 0 \nonumber .
\end{eqnarray}

Let us assume for a moment that none of these polynomials vanishes identically. Then, we can factorize 
each of them into irreducible polynomials 
\begin{equation}
\nonumber
p^{(k)}(z_k,U)=\Pi_{j=1}^l p^{(k)}_l(x_k,U).
\end{equation}
Every solution will be given by roots of some combination of $p^{(k)}_l$. Since there are finitely 
many combinations, and infinite solutions, there will be one particular combination that will give 
us solutions for all but finitely many values of $U$ (we are again using Lemma 4.1).

If one of these polynomials vanishes identically, it means that our system of equations is degenerate, 
and we can add equations to it, reducing the set of solutions, in such a way that the previous argument 
will work for the extended set of polynomial equations. 

Therefore, each of our variables $z_k$ will be an algebraic function of $U$ defined by
\begin{equation}
\label{algebraic}
p^{(k)}_l(z_k,U)=0.
\end{equation}

Since we know that our wavefunction is a rational function in $z_k$ and $U$, if we go from $U=\infty$ to 
any finite $U>0$ along the set of solutions defined by \eqref{algebraic}, our wavefunction will have finitely 
many zeros and poles. Since those are isolated, we can redefine the wavefunction at these points by 
the limit of the normalized wavefunctions as we approach them, since the energy of those states varies continuously and is 
equal to the ground state energy for all $U$. Again, this idea goes back to \cite{YY}, 
in the context of a spin system. That concludes the proof that the Bethe Ansatz gives us the true ground 
state of the system for any $U>0$.

If $N/N_a \leq 2/3$, we could actually prove a slightly stronger version of this result, in which we show that the 
wavefunction that results from the solution of \eqref{bethe1} and \eqref{bethe2} is not zero for all $U>0$.

\begin{theorem}
If the density of electrons per site satisfies
\begin{equation}
\nonumber
\frac{N}{N_a} \leq \frac{2}{3},
\end{equation}
the norm of the state given by the Bethe-Ansatz method is strictly positive:
\begin{equation}
\nonumber
\| \sum_{1\leq x_i \leq N_a}f(x_1,\dots ,x_N)|x_1,\dots ,x_N> \| >0.
\end{equation}
Therefore, the Bethe-Ansatz gives us the true and unique ground state of the system.
\end{theorem}

{\bf{Proof}}: Let us start by considering the case $M=M'=N/2$. 
Since we know that for a given solution of \eqref{bethe1} and \eqref{bethe2}, the coefficients $A_Q\{P\}$ 
are not 
all equal to zero, we just need to show that $f(x_1,\dots,x_N) \equiv 0 $ in the region $R$ implies 
\begin{equation}
\nonumber
A_Q\{ P\}=0, 
\end{equation}
for any $P$ and $Q$.

Let us consider the region $R_0$ defined by 
\begin{equation}
\nonumber
R_0=\{1 \leq x_{1} \leq x_{2} \leq \dots \leq x_{N} \leq N_a \} \subset R.
\end{equation}

Gaudin \cite{G} showed that $f(x_1,\dots,x_N)$ in $R$ is totally determined by its value 
in the region $R_0$, and $A_0\{ P \}=0$, for any P, implies $A_Q\{P\}=0$, for all $P,Q$. Therefore 
we just need to show that $f(x_1,\dots,x_N)\equiv 0$ implies $A_0\{ P \}=0$, for any P. We will denote 
the restriction of $f(x_1,\dots,x_N)$ to $R_0$ by 
\begin{equation}
\nonumber
f_0(x_1,\dots,x_N)=\sum_P A_0\{P\}\exp{\bigl[ i\sum_{j=1}^Nk_{P_j}x_j \bigr]}.
\end{equation}

Also in \cite{G}, it is shown that for any two permutations $P$ and $Q$, such that $Q$ only acts 
on variables corresponding to one kind of spin, we have
\begin{equation}
\nonumber
A_0\{PQ\}= I(Q) A_0\{P\},
\end{equation}
where $I(Q)$ is the sign of the permutation $Q$. 
This can be proved by combining the boundary conditions at the different regions $R_Q$ with the antisimmetry 
of the wavefunction with respect to exchange of identical particles.
In particular, if $P$ and $P'$ differ only by the exchange of coordinates of two identical particles, 
\begin{equation}
\nonumber
A_0\{P'\}= - A_0\{P\}
\end{equation}

Therefore, we have
\begin{eqnarray}
f_0(x_2,x_1,\dots,x_N)=\sum_P A_0\{P\}\exp{\bigl[ i(k_{P1}x_2+ k_{P2}x_1+ \cdots 
+k_{PN}x_N)\bigr]}= \nonumber \\
-\sum_{P'} A_0\{P'\}\exp{\bigl[ i(k_{P'1}x_1+ k_{P'2}x_2+ \cdots 
+k_{PN}x_N)\bigr]}=-f_0(x_1,x_2,\dots,x_N). \nonumber
\end{eqnarray}

Notice that this is not a trivial consequence of the antisymmetry alone, since $f_0$ is not
 the true wavefunction when $x_2 < x_1$.

As a result, we have that $f(x_1,\dots,x_N)\equiv 0$ implies $f_0(x_1,\dots,x_N)= 0$ in the 
region
\begin{equation}
\nonumber
\{1 \leq x_{1}, \dots, x_{M} \leq x_{M+1} , \dots, x_{N} \leq N_a \} \subset R
\end{equation}

If $N_a \geq 3N/2$, that includes
\begin{equation}
\nonumber
S=\{ 1 \leq x_{1}, \dots, x_{M} \leq N \, , \, N+1 \leq x_{M+1} , \dots, x_{N} \leq 3N/2 \} \subset R
\end{equation}

If $f_0$ is zero in $S$, we can fix $x_2, \dots, x_N$ and consider the system of equations
\begin{equation}
\nonumber
f_0(x_1, x_2, \dots ,x_N)=0 \quad , \quad 1 \leq x_1 \leq N,
\end{equation}
or
\begin{equation}
\label{linear}
e^{ik_1x_1}C^{(1)}(x_2,\dots, x_N)+\cdots + e^{ik_Nx_1}C^{(N)}(x_2,\dots, x_N)=0 \quad , 
\quad 1 \leq x_1 \leq N,
\end{equation}
where each $C^{(n)}$ is given by a sum over permutations such that $P(1)=n$:
\begin{equation}
\nonumber
C^{(n)}=\sum_{P:P1=n}A_0\{P\}\exp{\bigl[ i( k_{P2}x_2+ \cdots 
+k_{PN}x_N)\bigr]}.
\end{equation}

Therefore, \eqref{linear} is a homogeneous system of $N$ linear equations 
in the variables $C^{(N)}$. The determinant
\begin{equation}
\nonumber
\begin{pmatrix}
e^{ik_1} & e^{ik_2} & \cdots & e^{ik_N} \\
e^{2ik_1} & e^{2ik_2} & \cdots & e^{2ik_N} \\
\vdots & \vdots & \ddots & \vdots \\
e^{Nik_1} & e^{Nik_2} & \cdots & e^{Nik_N} 
\end{pmatrix}
=\prod_{j<i}(e^{ik_i}-e^{ik_j})\neq 0
\end{equation}
does not vanish, since $-\pi < k_1 < \dots < k_N <\pi$, and  we have
\begin{equation}
\nonumber
C^{(n)}=0 \quad , \quad 1 \leq n \leq N.
\end{equation}

We can fix $x_3, \dots, x_N$, and write
\begin{equation}
\nonumber
C^{(n)}=\sum_{m\neq n}e^{ik_mx_2}C^{(n,m)}(x_3,\dots, x_N)=0
\quad , \quad 1 \leq x_2 \leq N-1,
\end{equation}
where
\begin{equation}
\nonumber
C^{(n,m)}=\sum_{P:P1=n,P2=m}A_0\{P\}\exp{\bigl[ i( k_{P3}x_3+ \cdots 
+k_{PN}x_N)\bigr]}.
\end{equation}
Again, the determinant does not vanish and 
\begin{equation}
\nonumber
C^{(n,m)}=0.
\end{equation}
After repeating this argument for all spin-down variables, we will have only 
$M=N/2$ pseudo-momenta $k_i$ in each term. That is why it suffices to have $N+1 \leq x_{M+1} 
, \dots, x_{N} \leq 3N/2$. By the time we get to the variable $x_N$, we will have
\begin{equation}
\nonumber
A_0\{P\}=0
\end{equation}
for all $P$. 

If the set $\{k_i\}$ is given by a solution of the equations \eqref{bethe1} and \eqref{bethe2}, 
they correspond to a non trivial set of coefficients $A_0\{P\}$, and by the result above, 
$f(x_1,\dots ,x_N) \neq 0$ for some $(x_1, \dots, x_N ) \in R$, and the state given by the right 
side of \eqref{state} has a strictly positive norm. That concludes the proof of the theorem.

In general, $M\neq M'$, and the argument above still applies provided $N_a \geq N+M$ (which includes 
$N_a \geq 3N/2$ as a particular case).

\section{The thermodynamic limit}

Let us now analyze the behavior of the solutions $\{k_j,\Lambda_\alpha\}$ as we take the thermodynamic 
limit, by taking $N_a\rightarrow \infty$, keeping the ratios $n=N/N_a$ and $m=M/N_a$ fixed. 

We can define
\begin{equation}
\rho_i(k)=
\frac{1}{N_a|k_{j+1}-k_j|} \quad , \quad k_j \leq k < k_{j+1} ,
\end{equation}
and $\rho_i(k)=0$ for $k < k_1$ or $k\geq k_N$. The above expression defines a sequence of functions 
$\rho_i \in L^1([-\pi,\pi])$. As $i$ increases, so does the number of sites, and $i\rightarrow \infty$ 
corresponds to $N_a \rightarrow \infty$. Analogously, we can define $\sigma_i(\Lambda) \in L^1({\mathbb{R}})$ by 
\begin{equation}
\sigma_i(\Lambda)=
\frac{1}{N_a|\Lambda_{\beta+1}-\Lambda_\beta|} \quad , \quad \Lambda_\beta \leq \Lambda < \Lambda_{\beta+1} ,
\end{equation}
and $\sigma_i(\Lambda)=0$ for $\Lambda < \Lambda_1$ or $\Lambda \geq \Lambda_M$. From the definition, the $L^1$ norm of 
$\rho_i$ and $\sigma_i$ satisfy 
\begin{equation}
\| \rho_i \|_{L^1}=\frac{N-1}{N_a}=n-\frac{1}{N_a} \quad , \quad \| \sigma_i \|_{L^1}=\frac{M-1}{N_a}=m-\frac{1}{N_a}.
\nonumber
\end{equation}
In particular, these functions are uniformly bounded as we take the thermodynamic limit.

Let us also define
\begin{equation}
\label{deff}
f_i(k)=N_ak-\sum_\beta \theta(2\sin{k}-2\Lambda_\beta).
\end{equation}
From the Lieb-Wu equations, we have $f_i(k_j)=2\pi I_j$. Also,
\begin{equation}
\nonumber
f_i'(k)\leq N_a+\frac{8}{U}M \leq N_a \Bigl(1+\frac{8m}{U}\Bigr),
\end{equation}
which implies
\begin{equation}
f_i(k_{j+1})-f_i(k_j)=2\pi \leq |k_{j+1}-k_j|N_a \Bigl(1+\frac{8m}{U}\Bigr),
\nonumber
\end{equation}
or
\begin{equation}
\rho_i(k)\leq \frac{(1+8m/U)}{2\pi}.
\nonumber
\end{equation}
Since $\rho_i(k)$ is bounded, $\rho_i \in L^p([-\pi,\pi])$, for $1 \leq p \leq \infty$. The sequence 
$\{ \rho_i \}$ is actually uniformly bounded in $L^p$.

With a similar goal, we define
\begin{equation}
\nonumber
g_i(\Lambda)=\sum_j \theta(2\sin{k_j}-2\Lambda) +\sum_\beta \theta(\Lambda-\Lambda_\beta),
\end{equation}
such that $g_i(\Lambda_\alpha)=2\pi J_\alpha$. Its derivative is bounded by
\begin{equation}
g_i'(\Lambda) \leq \frac{8}{U}\sum_j\frac{1}{1+\bigl(\frac{4\sin{k_j}-4\Lambda}{U}\bigr)^2}
\leq \frac{8N_a n}{U}.
\label{gfunction}
\end{equation}
Once more, that implies
\begin{equation}
g_i(\Lambda_{\alpha+1})-g_i(\Lambda_\alpha)=2\pi \leq |\Lambda_{\alpha+1}-\Lambda_\alpha| N_a \frac{8n}{U},
\nonumber
\end{equation}
or
\begin{equation}
\sigma_i(\Lambda) \leq \frac{4n}{\pi U}.
\nonumber
\end{equation}
The first inequality on \eqref{gfunction} shows that all $g_i'$ are bounded by a function in $L^2(\mathbb{R})$. Therefore, 
$\{ \sigma_i \}$ is uniformly bounded in $L^2$.

Since both $\{ \rho_i \}$ and $\{ \sigma_i \}$ are uniformly bounded, we can use Banach-Alaoglu theorem to show that 
there is a subsequence 
converging weakly in $L^2$ (see \cite{LL}):
\begin{equation}
\rho_{n_i} \rightharpoonup \rho_w \quad , \quad \sigma_{n_i} \rightharpoonup \sigma_w.
\nonumber
\end{equation}

We will be working now with this subsequence, but we will keep the index as $i$ just to simplify the 
notation. Our goal is to show that this subsequence also converges pointwise.

If we take equation \eqref{bethe1} with index $j$, subtract it from the same equation with index 
$j+1$, and divide the result by $2\pi N_a|k_{j+1}-k_{j}|$, we obtain
\begin{equation}
\label{discretek}
\rho_i(k)=\frac{1}{2\pi}-\frac{1}{2\pi}\sum_\beta \frac{\theta(2\sin{k_{j+1}}-2\Lambda_\beta)-
\theta(2\sin{k_j}-2\Lambda_\beta)}{k_{j+1}-k_j}\sigma_i(\Lambda)|\Lambda_{\beta+1}-\Lambda_\beta|,
\end{equation}
where $k_j \leq k < k_{j+1}$ and $\Lambda_\beta \leq \Lambda < \Lambda_{\beta+1}$. We can fix $k$ now 
and consider \eqref{discretek} as an equation for $\rho_i(k)$.

To show the existence of a pointwise limit, we just need to show that the right side of \eqref{discretek} 
converges for all $k$. We know that the fraction on the right side converges to the derivative of $\theta$, if 
$|k_{j+1}-k_j|\rightarrow 0$. Since we did not prove that this happens for every $k$, let us divide the domain of 
$\rho$ into $Q\in [-\pi,\pi]$, defined by $Q=\{k \in [-\pi,\pi]:|k_{j+1}-k_j| \rightarrow 0\}$, and its complimentary $Q^c$. 

In $Q^c$, the distance between the surrounding $k_j$ does not converge to zero. For a $k$ in the interior of $Q^c$,
we can find an $\epsilon$ and a 
subsequence such that $|k_{j}-k|> \epsilon/2$, for all $j$. That implies that the weak limit vanishes in this interval 
of lenght $\epsilon$ around $k$, since the integral of the characteristic function of any subset of this interval 
times $\rho_i$ converges to zero.  
Also, the subsequence will have a pointwise limit on this set that is equal to zero (almost everywhere). Since the boundary of 
$Q^c$ has measure zero, the value of the function at those points is not relevant in the context of 
$L^p$-spaces, which are defined by equivalence classes of functions that differ only on a set of measure zero. 


Therefore, the subsequence $\{ \rho_i\}$ has a pointwise limit in $Q^c$ that agrees with the weak limit. 
We will show that the same holds true in $Q$. A similar analysis can be done for the sequence $\{ \sigma_i\}$. 
We define the set $B\in \mathbb{R}$ by $B=\{\Lambda \in \mathbb{R}:|\Lambda_{\alpha+1}-\Lambda_\alpha| \rightarrow 0\}$. In 
its complimentary $B^c$, the subsequence has a zero pointwise limit that agrees with the weak limit.

Now if $k \in Q$, the last term of \eqref{discretek} is
\begin{equation}
\label{firstaprox}
\frac{\cos{k}}{\pi}\sum_\beta [\theta'(2\sin{k}-2\Lambda_\beta)+E^{(1)}_i(k,\Lambda_\beta)]
\sigma_i(\Lambda)|\Lambda_{\beta+1}-\Lambda_\beta|,
\end{equation}
where $E^{(1)}_i(k,\Lambda_\beta)$ is the error in approximating the ratio in \eqref{discretek} 
by the derivative of $\theta$. Since $\theta''$ is bounded, we have
\begin{equation}
\nonumber
\sum_\beta E^{(1)}_i(k,\Lambda_\beta)
\sigma_i(\Lambda)|\Lambda_{\beta+1}-\Lambda_\beta| \leq \sum_\beta
\frac{C|k_{j+1}-k_{j}|}{N_a}=Cm|k_{j+1}-k_{j}| \longrightarrow 0,
\end{equation}
as $i\rightarrow \infty$.

We need to show now that \eqref{firstaprox} converges to an integral. We first notice that 
\begin{eqnarray}
\sum_\beta \theta'(2\sin{k}-2\Lambda_\beta)\sigma_i(\Lambda)|\Lambda_{\beta+1}-\Lambda_\beta|= 
\int \theta'(2\sin{k}-2\Lambda_\beta)\sigma_i(\Lambda) d\Lambda + \nonumber \\
+\int [\theta'_{tr}(2\sin{k}-2\Lambda_\beta)-\theta'(2\sin{k}-2\Lambda_\beta)]\sigma_i(\Lambda) d\Lambda,
\label{secondaprox}
\end{eqnarray}
where $\theta'_{tr}$ is the truncated version of $\theta'$, defined by a piecewise 
constant function, whose value on the interval 
$[\Lambda_\beta,\Lambda_{\beta+1})$ is equal to $\theta'(2\sin{k}-2\Lambda_\beta)$. Again, 
since the second derivative of $\theta$ is bounded uniformly by a function in $L^1$, we can write
\begin{equation}
\nonumber
|\theta'_{tr}(2\sin{k}-2\Lambda_\beta)-\theta'(2\sin{k}-2\Lambda_\beta)| \leq g(\Lambda)|\Lambda_{\beta+1}-\Lambda_\beta|,
\end{equation}
where $g(\Lambda) \in L^1(\mathbb{R})$. It follows that
\begin{equation}
\lim_{i\rightarrow \infty}\int |\theta'_{tr}(2\sin{k}-2\Lambda_\beta)-\theta'(2\sin{k}-2\Lambda_\beta)|\sigma_i(\Lambda) d\Lambda
 \leq \lim_{i\rightarrow \infty} \int \frac{g(\Lambda)}{N_a}d\Lambda =0. \nonumber
 \end{equation}

Since $\{\sigma_i \}$ has a weak limit, the first integral on the right side of \eqref{secondaprox} converges, 
and we have
\begin{equation}
\lim_{i\rightarrow \infty}\rho_i(k)=\frac{1}{2\pi}-\frac{\cos{k}}{\pi}\int_B\theta'(2\sin{k}-2\Lambda)\sigma_w(\Lambda) d\Lambda,
\end{equation}
for all $k \in Q$. A similar analysis for $\sigma_i(\Lambda)$ yields
\begin{equation}
\lim_{i\rightarrow \infty}\sigma_i(\Lambda)=-\frac{1}{\pi}\int_B\theta'(2\sin{k}-2\Lambda)\rho_w(\Lambda) d\Lambda+
\frac{1}{2\pi}\int_B\theta'(\Lambda-\Lambda')\sigma_w(\Lambda) d\Lambda,
\end{equation}
for all $\Lambda \in B$. 

We just showed that the subsequences have a pointwise limit. But since $\{ \rho_i\},\{\sigma_i\}$ are uniformly bounded, 
dominated convergence implied the existence of a strong limit, which equals the pointwise limit and the weak limit almost everywhere. 
Therefore, as we go to the thermodynamic limit on this subsequence, our $k$-s and $\Lambda$-s converge to 
distributions satisfying the integral equations 
\begin{eqnarray}
\label{integralI}
\rho(k) & = & \frac{1}{2\pi}+\cos{k}\int_BK(\sin{k}-\Lambda)\sigma(\Lambda) d\Lambda, \\
\sigma(\Lambda) & = & \int_QK(\sin{k}-\Lambda)\rho(k) dk-
\int_BK^2(\Lambda-\Lambda')\sigma(\Lambda) d\Lambda,
\label{integralII}
\end{eqnarray}
where we define 
\begin{eqnarray}
K(\Lambda-\Lambda') & = & -\frac{1}{\pi}\theta'(2\Lambda-2\Lambda'), \nonumber \\
K^2(\Lambda-\Lambda') & = & -\frac{1}{2\pi}\theta'(\Lambda-\Lambda')=\int_{-\infty}^{\infty}
K(\Lambda-x)K(x-\Lambda')dx. \nonumber
\end{eqnarray}

It was shown in \cite{LW2} that it is convenient to restrict the integral in $k$ to a domain where 
the function $\sin{k}$ is monotonic. Hence, we divide the range of integration $Q$ into 
$Q_-=Q\cap[-\pi/2,\pi/2]$ and $Q_+=Q\cap([-\pi,-\pi/2)\cup(\pi/2,\pi])$. The first thing to show is that 
for every $k=(\pi/2+x) \in Q_+$, there is a $k=(\pi/2-x) \in Q_-$. 

Since the derivative of $f_i(k)$, as defined in \eqref{deff} satisfies $f'_i(k)\geq N_a$ for $k \in [-\pi/2,\pi/2]$, the set 
$Q_-$ consists of a single interval $[-k^*,k^*]$ (possibly $[-\pi/2,\pi/2]$). Since $f(k) \geq f(k^*)+N_a(k-k^*)$ for 
any $k \in [k^*,\pi-k^*]$, and the $k_j$-s are ordered, we conclude that the existence of a gap 
in the interval $[k^*,\pi/2]$ implies that $Q_+\cap [\pi/2,\pi-k^*]=\emptyset$, and for every positive $k$ in $Q_+$ there 
is a corresponding point in $Q_-$ given by the reflection with respect to $k=\pi/2$. The same holds for for every negative 
$k$ in $Q_+$, where now we need to take the reflection with respect to $-\pi/2$.

Therefore, if we define the set $Q_+^r$ by the reflection of $Q_+$ with respect to $\pm \pi/2$,
\begin{equation}
Q_+^r=\{\pm (\pi-k) \, , \, k \in Q_+\cap [\pi/2,\pi]\},
\nonumber
\end{equation}
and we will have $Q_+^r \subset Q_-$. Finally, we can avoid the integration in $Q_+$ by noting that 
$K(\sin{k}-\Lambda)$ is an even function of $k-\pi/2$, which implies by \eqref{integralI} that 
$\rho(k)-1/2\pi$ is odd as a function of $k-\pi/2$. Therefore, we can rewrite part of the first integral 
in \eqref{integralII} as
\begin{equation}
\nonumber
\int_{Q_+}+\int_{Q_+^r}K(\sin{k}-\Lambda)\rho(k) dk=\int_{Q_+}+\int_{Q_+^r}\frac{K(\sin{k}-\Lambda)}
{2\pi}=\frac{2}{2\pi}\int_{Q_+^r}K(\sin{k}-\Lambda)
\end{equation}

Now $\sin{k}$ is monotonic in the domain of the integral in $k$, and we can make the change in variables
\begin{eqnarray}
t(x) & = & \frac{(1-x^2)^{-1/2}}{2\pi}, \nonumber \\
f(x) & = & (1-x^2)^{-1/2}\rho(\sin^{-1}{x}), \nonumber
\end{eqnarray}
for $-1 \leq x \leq 1$. If $A,D$ are the images of $Q_-,Q_+^r$ under $\sin{k}$, the integral equations \eqref{integralI}, \eqref{integralII} can now be written
\begin{eqnarray}
\label{integralf}
f(x) & = & t(x)+\int_{-\infty}^{\infty}K(x-x')B(x')\sigma(x')dx' \quad , \quad x \in A, \\
\sigma(x) & = &\int_{-\infty}^{\infty}K(x-x')(A(x')f(x')+D(x')t(x'))dx' \nonumber \\
\label{integralsigma}
& & -\int_{-\infty}^{\infty}K^2(x-x')B(x')\sigma(x')dx' \quad , \quad x \in B,
\end{eqnarray}
where $A(x)=\chi_{A \setminus D}$, $B(x)=\chi_B$, and $D(x)=2\chi_{D}$ ($\chi_X$ denotes the 
characteristic function of the set $X$).

These equations have to be solved for $x$ in the sets defined above. However, we can let the right side 
of \eqref{integralf} and \eqref{integralsigma} define $f(x)$ and $\sigma(x)$ for any $x \in \mathbb{R}$, 
by setting $t(x)\equiv 0$ for $|x|>1$.
Since 
the integrals only depend on $f(x)$ and $\sigma(x)$ at the original intervals, there is a one to one correpondence 
between the solutions to the extended equations and the solutions of the original equations. In particular, 
the new equations will have a unique solution if and only if the solution to \eqref{integralf} and \eqref{integralsigma} 
is unique.

It is convenient to write the integral equations in the operator form
\begin{eqnarray}
\label{operatorI}
f & = & t+\hat{K}\hat{B}\sigma, \\
\label{operatorII}
\sigma & = & \hat{K}\hat{A}f+\hat{K}\hat{D}t-\hat{K}^2\hat{B}\sigma,
\end{eqnarray}
where $\hat{K}$ is the convolution with $K$ and $\hat{A}$, $\hat{B}$, $\hat{D}$ are the 
multiplication by $A$, $B$ and $D$.

The proof that \eqref{operatorI} and \eqref{operatorII} have a unique solution for given sets 
$Q$ and $B$ was done by Lieb and Wu in \cite{LW2}. We did not prove here that the sets $Q$ and $B$ are uniquely 
defined for every $n=N/N_a$ and $m=M/N_a$, although we believe they are (and we also belive that they 
are given by intervals $Q=[-k_{max},k_{max}]$, $B=[-\Lambda_{max},\Lambda_{max}]$). However, this is not 
a problem for the absolute ground state of the half filled band, in which case an explicit solution for 
\eqref{operatorI} and \eqref{operatorII} can be found. In this case, we have $n=1$ and $m=1/2$, 
and Lieb and Wu proved that $Q=[-\pi,\pi]$ and $B=\mathbb{R}$. Let us restrict now to this case.

Since the solution is unique, and every subsequence of $\{\rho_i \}$, $\{ \sigma_i \}$ contains a further subsequence that 
converges to the same $\rho$ and $\sigma$, we can actually prove that the whole sequence converges. Otherwise, we would be able 
to find a subsequence such that $\|\rho_{n_i}-\rho \|>\epsilon$ or $\|\sigma_{n_i}-\sigma \|>\epsilon$, which is a contradiction. 

The limit distribution in the thermodynamic limit for $k$-s and $\Lambda$-s are then given 
by the solution of \eqref{integralI}, \eqref{integralII}, obtained in \cite{LW1},
\begin{eqnarray}
\label{solI}
\rho(k)& = & \frac{1}{2\pi}+\frac{\cos{k}}{\pi}\int_0^\infty \frac{\cos{(\omega \sin{k})}
J_0(\omega)}{1+e^{\omega U/2}}d\omega, \\
\label{solII}
\sigma(\Lambda) & = & \frac{1}{2\pi}\int_0^\infty \frac{J_0(\omega)\cos{(\omega \Lambda)}}
{\cosh{(\omega U/4)}}d\omega,
\end{eqnarray}
which yield the ground state energy
\begin{equation}
\nonumber
E=-2N_a \int_{-\pi}^\pi \rho(k) \cos{k} dk=-4N_a\int_0^\infty \frac{J_0(\omega)J_1(\omega)}{\omega
(1+e^{\omega U/2})}d\omega,
\end{equation}
where $J_n$ is the Bessel function of order $n$.

\section{Conclusions}

We have concluded the proof that the Bethe Ansatz equations, and in particular the Lieb-Wu equations 
\eqref{bethe1} and \eqref{bethe2}, have a solution for finite $M$, $N$ and $N_a$. The solution is indeed real 
and ordered in the indices $j$ and $\alpha$, as assumed by Lieb and Wu in the derivation of the integral 
equations for the model in the thermodynamic limit.

We have also shown the existence of a continuous set of solutions extending from any $U>0$ to $U=\infty$.
The continuity is important to show that the state obtained by the Lieb-Wu equations is indeed the 
ground state of the system, provided it does not vanish. Since we know that the state obtained is the genuine ground state of the system 
at $U=\infty$, and that the ground state is nondegenerate for any $U>0$ (for $M,M'$ odd),  
a normalizable state given by the Bethe Ansatz should also give us the ground state of the system for any $U$, 
since the energy cannot discontinuously jump from an excited level to the ground state as we increase 
the interaction strength from a finite value to the limit $U=\infty$. 

To conclude the proof for a finite lattice, we 
show that the norm of the wavefunction obtained by the Bethe Ansatz equations is well defined and not zero, for all $U>0$, except for finitely many values 
of $U$. If $U_0$ is one of this values, the correct eigenfunction will be given by the limit $U\rightarrow U_0$ of the normalized wavefunction, 
since $U_0$ is an isolated zero or singularity.

Going from the finite lattice to the thermodynamic limit, we focus primarily on the half-filled band $N=N_a$. In 
this case, the set of $k$-s converge to a distribution $\rho(k)$ on the interval $[-\pi,\pi]$. If we consider the 
absolute ground state with $M=M'$
, the set of $\Lambda$-s converge to a distribution $\sigma(\Lambda)$ in $\mathbb{R}$, and we showed that 
$\rho(k)$ and $\sigma(\Lambda)$ satisfy the integral equations derived by Lieb and Wu, whose 
solution is given by \eqref{solI}, \eqref{solII}. For other fillings,
we also prove the convergence (at least for a subsequence) of the distribution to the solution of the integral 
equations, but we do not prove uniqueness of the sets $Q$ and $B$ that contain the support of $\rho$ and 
$\sigma$.

The author is indebted to Elliott H. Lieb for useful comments and suggestions.

\appendix

\section{Groebner bases}

Every ideal $I(f_1,\dots,f_m)$ generated by the polynomials $\{f_k\}_{k=1}^m\neq \{0\}$ has a 
Groebner basis. The general idea in constructing a Groebner basis is to eliminate one by one 
the variables of the polynomials $f_k$, obtaining a set of 
polynomials $G=\{g_k\}_{k=1}^l$ that generate the same ideal as $\{f_k\}_{k=1}^N$. One of its main applications is to solve systems of polynomial 
equations, since reducing the number of variables simplifies the task of finding solutions. 
This process can be thought of as a generalization of the process of solving 
a system of linear equations by elimination of variables. We will present here the essential facts about 
Groebner basis in order to understand the results of section 4. The interested reader should refer to \cite{CLO} 
for a precise definition of a Groebner basis and the algorithm used to obtain it.

Given an ideal $I \subset \mathbb{C}[x_1,\dots, x_n]$, we can define the $k$-elimination ideal by
\begin{equation}
\nonumber
I_k=I\cap \mathbb{C}[x_{k+1},\dots,x_n].
\end{equation}
$I_k$ is the set of all polynomials in $I$ that do not depend on the first $k$ variables $x_k$. What makes 
the Groebner basis special is that 
\begin{equation}
G_k=G\cap \mathbb{C}[x_{k+1},\dots,x_n]
\nonumber
\end{equation}
is also a basis for $I_k$. 

As a consequence of this property, if the variety defined by \eqref{polequation} consists of 
finitely many points, $G_{n-1}\neq\{0\}$, and we can solve the corresponding equations for $x_n$, to find 
$V(I_{n-1})$. Then, we can move on to $G_{n-2}$, and solve it for $x_{n-2}$, finding $V(I_{n-2})$, and so 
on. In the end, we will find $V(I)$ just by solving polynomial equations in one variable. We are ordering the variables here by $x_1>x_2>\dots >x_n$. This ordering determines 
which variables will be eliminated first. Of course we can choose any other ordering, and we will 
obtain a different Groebner basis, but the set of solutions should be the same

As one example, consider the set of polynomial equations $f_1=x^2 + y^2 + z - 1=0$, $f_2=x + y + z - 1=0$ 
and $f_3=x + y - z^2 - 1=0$. 
Fixing the ordering $x>y>z$, the ideal $I(f_1,f_2,f_3)$ has $G=\{x+y+z-1,y^2-y+yz-z,z^2+z\}$ as a possible Groebner basis. Therefore, the solution 
to the original equations will satisfy
\begin{eqnarray}
x+y+z-1=0, \nonumber \\
y^2-y+yz-z=0, \nonumber \\
z^2+z=0, \nonumber
\end{eqnarray}
which can be easily solved to get $V(I)=\{(0,1,0),(1,0,0),(1,1,-1)\}$.

Clearly our solutions $V(I)$ could be more complicated than isolated points. It could contain subsets of higher 
dimension. In this case, we would have a slightly different basis. Consider for instance the set of solutions 
of $f_1=f_2=0$. It clearly has infinitely many solutions, and $G=\{x+y+z-1,2y^2-2y+2yz+z^2-z\}$ is a Groebner basis. We see 
that $G_2=\{0\}$ and $V(I_2)=\mathbb{C}$. In this case, we could not eliminate $x$ and $y$, since we have solutions 
for infinitely many values of $z$. In summary, the number of variables we can actually eliminate by 
calculating the Groebner basis depends on the dimension of $V(I)$.

\end{document}